\documentclass[12pt]{article}
\usepackage{graphicx}
\usepackage{epstopdf}
\usepackage{subfigure}

\usepackage{epsfig,amsmath,amssymb,verbatim,mathrsfs}

\def\beq{\begin{eqnarray}}
\def\eeq{\end{eqnarray}}
\def\bea{\begin{eqnarray}}
\def\eea{\end{eqnarray}}

\newcommand{\be}{\begin{equation}}
\newcommand{\ee}{\end{equation}}

\begin{document}

\setlength{\baselineskip}{0.2in}


\begin{titlepage}
\noindent
\flushright{March 2013}
\vspace{0.2cm}

\begin{center}
  \begin{Large}
    \begin{bf}
The Simplest Models of Radiative Neutrino Mass: Excluding Simplified Zee Models and Beyond\\

     \end{bf}
  \end{Large}
\end{center}

\vspace{0.2cm}

\begin{center}

\begin{large}

{Sandy~S.~C.~Law$^{*}$\footnote{slaw@mail.ncku.edu.tw}  and Kristian~L.~McDonald$^{\dagger}$\footnote{klmcd@physics.usyd.edu.au}}\\
     \end{large}
\vspace{0.5cm}
  \begin{it}
* Department of Physics, National Cheng-Kung University, \\ Tainan 701, Taiwan\\
\vspace{0.5cm}
$\dagger$ ARC Centre of Excellence for Particle Physics at the Terascale,\\
School of Physics, The University of Sydney, NSW 2006, Australia\\\vspace{0.5cm}
\vspace{0.3cm}
\end{it}
\vspace{0.5cm}

\end{center}


\begin{abstract}
The complexity of radiative neutrino-mass models can be judged by: (i) whether they require the imposition of \emph{ad hoc} symmetries, (ii) the number of new multiplets they introduce, and (iii) the number of arbitrary parameters that appear. Adopting the view that the imposition of arbitrary new symmetries is the least appealing approach, the simplest models have two new multiplets and a minimal number of new parameters. With this in mind, we search for the simplest models of radiative neutrino mass. We are lead to two new models, containing a real scalar triplet and a charged scalar doublet (respectively), in addition to the charged singlet scalar considered by Zee [$h^+\sim(1,1,2)$]. The new models are essentially simplified versions of the Zee model and appear to be \emph{the simplest} models of radiative neutrino mass. However, these models are only of pedagogical interest; despite successfully generating nonzero masses, present-day data is sufficient to rule them out. The lessons learnt from these models also enable one to exclude a more general class of radiative models. Moving beyond the minimal cases, we find a new model of two-loop masses that employs the charged doublet $\Phi\sim (1,2,3)$ and the doubly-charged scalar $k^{++}\sim(1,1,4)$.

\end{abstract}

\vspace{1cm}

\end{titlepage}

\setcounter{page}{1}


\vfill\eject


\section{Introduction\label{sec:introduction}}
If Nature employs a high UV scale to achieve small neutrino masses via a seesaw, we are unlikely to directly probe the origin of neutrino mass. In this case, circumstantial or indirect evidence may be our best guide. Such evidence can be acquired by systematically ruling out alternative low-energy approaches. Thus, it is useful to eliminate alternative models, where possible, by demonstrating their falsity.

Radiative models of neutrino mass~\cite{Zee:1980ai,Zee:1985id,Babu:1988ki} offer an attractive means of generating small neutrino masses, precisely because they afford a greater likelihood of being experimentally probed. There are, however, many candidate models in the literature. These models can be classified by various measures, including, e.g., the mass-dimension of the low-energy effective operator they induce; see e.g.~\cite{Bonnet:2012kz,Angel:2012ug}. Such classifications can be helpful to ensure a systematic exploration of the viable possibilities. 

One obvious quality to employ as a basis of classification is the simplicity, or complexity, of a given model. As in other fields of research, Occam's razor can be invoked to both classify existing models and guide the search for new possibilities. Ultimately, any  model must be judged by its concordance with Nature, regardless of whether or not it manifests an underlying simplicity.  None the less, scientific progress has shown that limiting the redundancy of our descriptions has often proved to be a fruitful approach.

In this work we aim to identify the simplest models of radiative neutrino mass. We provide criteria by which the complexity of a given model can be assessed, and use these measures to guide our search for the simplest models. 
In undertaking this task we arrive at two new radiative neutrino-mass models which may well be the simplest. The models are related the earliest proposal in the literature,  and can be thought of as simplified versions of the original Zee model~\cite{Zee:1980ai}.   

Beyond these minimal models, the Zee~\cite{Zee:1980ai} and Zee-Babu~\cite{Zee:1985id,Babu:1988ki} models appear to be unique in requiring just two additional fields to be added to the Standard Model (SM); that is, the only radiative mass models with just two beyond-SM multiplets, that are devoid of \emph{ad hoc} new symmetries, are the Zee model, the Zee-Babu model, and the two new models presented here. As we will see, however, there is a price to be paid for the simplicity of the new models  --- present-day data is sufficient to rule them out.  If one allows for additional complexity, there are a number of viable models that employ three beyond-SM fields. We present a new such example, achieving mass at the two-loop level with the aid of the charged doublet $\Phi\sim(1,2,3)$ and the doubly charged scalar $k^{++}\sim(1,1,4)$. In the course of our work we also uncover lessons that enable one to rule out a larger class of radiative mass models, purely on the basis of mass and mixing data. These observations truly  demonstrate the power of recent advances in experimental neutrino physics.

The plan of this paper is as follows. In Section~\ref{sec:criteria} we present criteria for the simplicity of radiative neutrino-mass models, and show how these measures can guide one in the search for new models. Section~\ref{sec:simple_zee} presents the variant Zee models, which we suspect to be the simplest possibilities. The search for minimal models continues in Section~\ref{sec:simple_zee_babu}, and new non-minimal models of radiative mass are described in Section~\ref{sec:more_complexity}. We conclude in Section~\ref{sec:conc}, and briefly discuss  a prior work in an Appendix.
\section{Assessing Complexity\label{sec:criteria}}
Neutrinos are massless in the renormalizable Standard Model. The experimental discovery of neutrino mass therefore provides concrete evidence for physics beyond this framework. At the non-renormalizable level, the SM contains the famous $d=5$ Weinberg operator $\mathcal{O}_W=(LH)^2/\Lambda$~\cite{Weinberg:1979sa}, which produces Majorana neutrino masses after electroweak symmetry breaking (here $H$ ($L$) is the SM scalar (lepton) doublet). It is encouraging that the minimal non-renormalizable extension of the SM automatically leads to massive neutrinos, and thus perfectly accommodates our concrete evidence for beyond SM physics. Unfortunately this tells us very little about the underlying origin of neutrino mass. To gain further insight into this matter one is forced to consider specific UV completions for this operator, or models that produce neutrino mass via a $d>5$ non-renormalizable operator in the low energy limit (see e.g.~\cite{Bonnet:2009ej}).

Radiative models of neutrino mass~\cite{Zee:1980ai,Zee:1985id,Babu:1988ki} provide a UV completion~\cite{Ma:1998dn} that has a higher probability of being falsifiable, simply because the new physics can be lighter --- perhaps even at the TeV scale. There are a large number of models with radiative neutrino mass in the literature, all of which require one to extend the SM to include new multiplets and additional free parameters. Very often these extensions also require that new symmetries be imposed, in order to ensure that non-desirable Lagrangian terms, and/or experimentally precluded processes, are absent. Given that these three ingredients form a common basis for candidate models of radiative neutrino mass, they can be used to classify the different models. In this work we use these ingredients to classify models on the basis of simplicity. We then use the criterion of minimality to arrive at simple new models.\footnote{Our discussion in this section is presented in some detail in order to clarify our motives and justify our conclusions. Readers disinterested in these details can find the resulting models in Sections~\ref{sec:simple_zee} and~\ref{sec:simple_zee_babu}.}

We propose that the simplicity of a given model of radiative neutrino mass can be judged by the following criteria:
\begin{itemize}
\item Whether the viability of the model relies on the imposition of \emph{ad hoc} new symmetries
\item The number of new (beyond SM) particle multiplets the model requires
\item The number of free parameters introduced 
\end{itemize}
Adopting the viewpoint that models whose success relies on the imposition of new symmetries are the least desirable, the simplest models are those with the least number of new multiplets, and the smallest number of additional parameters. 

The notion that models with arbitrary new symmetries are the least desirable is not the only view possible. As justification for this view, we note that the SM works successfully without  the imposition of any (otherwise arbitrary) global symmetries, be they discrete or continuous. Rare (or perhaps forbidden) processes, like proton decay or flavor changing effects, that can otherwise violate symmetries of the SM, are generated at the non-renormalizable level, and can be safely suppressed by high UV scales. This provides a simple explanation for the low-energy absence of such symmetry breaking effects. In retaining the spirit of the SM, one can therefore argue that a need to impose additional global symmetries to ensure viability is less attractive. 

Of course there may be additional gauge symmetries in Nature, whose presence in the UV influences the low energy manifestation of neutrino mass. From our current perspective, we note only that models with additional gauge symmetries in the UV do not appear to allow radiative models with less multiplets and free parameters. Thus, when judged purely from the perspective of simplicity, such models appear to offer no advantage. In pursuing the simplest models of radiative neutrino mass we thus exclude the possibility of imposing new symmetries. One may also wonder if a preference should be given to multiplets forming smaller gauge representations. This matter does not appear significant for multiplets in either the fundamental or adjoint representations, given that both of these representations are found in Nature. However, it may provide an additional measure of complexity in models with larger representations. Let us also emphasize that the mechanism of neutrino mass employed by Nature may or may not show an inclination towards simplicity. Still, we argue that the criterion of simplicity is a worthy perspective from which to consider the matter, absent evidence to the contrary. 

The minimal number of new multiplets required to generate radiative neutrino masses is two. This is seen as follows. Absent new symmetries, one does not expect minimal models to radiatively generate Dirac neutrino mass; the inclusion of a right-chiral singlet neutrino (or real triplet of fermions~\cite{Foot:1988aq}) would produce tree-level masses that cannot be excluded without new symmetries. Other fermion multiplets forming larger representations of $SU(2)_L$, that contain a right-chiral neutrino, also require new scalars in order to permit the necessary couplings.\footnote{The exception being the addition of vector-like triplet leptons, $E\sim(1,3,-2)$, which can minimally couple to the SM~\cite{Chua:2010me}. On their own, these do not admit massive SM neutrinos~\cite{Law:2012mj}.} Thus, we can focus on Majorana neutrinos.

Models with radiative Majorana masses must contain a source of lepton number violation,  requiring at least one term that explicitly breaks the lepton number symmetry of the renormalizable SM. The addition of a single new multiplet (call it $\mathcal{F}$) to the SM is insufficient to generate radiative Majorana masses. If $\mathcal{F}$ does not couple directly to SM fermions there is no way for the lepton number symmetry to be broken; in this case one has $L_\mathcal{F}=0$ and lepton number symmetry is conserved, irrespective of the quantum numbers of $\mathcal{F}$. 

If $\mathcal{F}$ couples directly to SM fermions, one can always define lepton number symmetry such that all terms involving SM fermions conserve the symmetry. This direct coupling is therefore insufficient, and if a single new multiplet is to do the job, it must also appear in the scalar potential in such a way that lepton number symmetry is broken. In this case the scalar potential is of the form $V=V(H,\mathcal{F})$. One can systematically explore the candidate scalar fields that have a renormalizable coupling to SM fermions and can break lepton number symmetry via the scalar potential. The only such candidate is the complex triplet $\Delta\sim(1,3,2)$, which admits the lepton number symmetry breaking term $V\supset \mu H\Delta H$. However, the presence of this linear term (in $\Delta$) also ensures that $\langle \Delta\rangle\ne0$, and thus generates tree-level neutrino masses via the allowed Yukawa coupling $\mathcal{L}\supset \lambda \bar{L^c}\Delta L$~\cite{type2_seesaw}. We can therefore exclude this possibility and conclude that it is not possible to generate radiative Majorana masses via the extension of the SM by a single new multiplet.\footnote{If, by some new mechanism, the direct Yukawa coupling is absent, the triplet can be employed in concert with a doubly charged singlet to generate mass at the two-loop level~\cite{Chen:2007dc}; also see Ref.~\cite{Kajiyama:2013zla}.}

Based on the above, and the fact that radiative models with two new multiplets are known to exist~\cite{Zee:1980ai,Zee:1985id,Babu:1988ki,Ma:1998dn}, one deduces that the minimum number of new multiplets necessary to generate radiative neutrino mass is two. 

Next we would like to determine which new multiplets are suitable to achieve the simplest constructs. A key matter is whether one should consider multiplets with nonzero $SU(3)_c$ charge or instead use $SU(3)_c$ singlets. First lets consider the use of colored multiplets. In this case one finds that the minimal number of new multiplets needed to achieve radiative neutrino masses is two~\cite{Ma:1998dn,Foot:2007ay}. For example, consider adding $h_1\sim(3,1,-2/3)$ and $h_2\sim(\bar{3},2,-1/3)$ to the SM. Then lepton number symmetry is broken explicitly by the term $\mu h_1h_2H\subset V(H,h_1,h_2)$ and Majorana neutrino masses appear at the one-loop level~\cite{Foot:2007ay}. However, a baryon number symmetry must be imposed to forbid rapid proton decay; a situation which appears to be generic,\footnote{See Ref.~\cite{Babu:2010vp} for another recent example.} and thus increases the complexity of models with two colored fields. 

Models with three new colored fields can successfully achieve neutrino masses at the one-loop~\cite{Ma:1998dn,FileviezPerez:2009ud,FileviezPerez:2010ch} and two-loop~\cite{Babu:2011vb} level.\footnote{We briefly comment on Ref.~\cite{FileviezPerez:2009ud} in the Appendix.} Furthermore, if the $SU(3)$ representation of the new multiplets is large enough, proton decay is not induced and no new symmetry is required~\cite{FileviezPerez:2009ud}. However, it is already well known that radiative models with two new fields exist, so clearly such models are not the simplest cases. Thus we can focus our attention on models with new fields that are $SU(3)_c$ singlets.

When seeking to add a (non-colored) source of lepton number violation to the SM, an obvious approach is to consider the lepton number violating fermionic bilinears. The alternatives are to (i) Yukawa couple a new real fermion to $L$ or $e_R$, such that a bare Majorana mass term also appears, which leads to tree-level masses, or (ii) add a non-real fermion, in which case multiple fermions are required to cancel anomalies, and at least a single new scalar is needed to ensure lepton number violation. We therefore focus on the lepton number violating bilinears. The list of particles that couple to such bilinears, constructed with SM leptons, is very small (see e.g.~\cite{Nieves:1981tv} or the second paper in~\cite{type2_seesaw}). From the direct product  $\overline{L^c}\otimes L$ one has the scalar triplet $\Delta\sim(1,3,2)$ and the singlet $h^+\sim(1,1,2)$. The former field is unsuitable for reasons already explained, the latter is employed in the Zee model~\cite{Zee:1980ai}. From the product $\overline{e^c_R}\otimes e_R$ one can also couple the scalar $k^{++}\sim(1,1,4)$, as employed in the Zee-Babu model~\cite{Zee:1985id,Babu:1988ki}. Based on their agreement with our criterion of simplicity thus far, it is clear to see why the Zee and Zee-Babu models were the earliest proposals in the literature! 

If we add either of the fields $h^+$ or $k^{++}$ to the SM with the goal of arriving at the simplest construct, we should next ask ``What is the appropriate second multiplet to ensure minimality?" This question requires one to consider the number of new free parameters that the resulting models contain. Although a literal counting of individual parameters is pedantic to a level beyond even our ambitions, there are some general rules that can be used to guide one's search. It is clear that new multiplets that can Yukawa-couple to SM fields are more likely to introduce a sizable number of new parameters, due to the existence of three generations. A new multiplet that cannot Yukawa-couple to the SM is therefore likely to require less free parameters. Thus, ideally the new multiplet will not Yukawa-couple to the SM. In addition the new multiplet should permit a term in the potential containing $h^+$ (or $k^{++}$) that explicitly breaks lepton number symmetry. With these demands established, one can systematically study the scalar potentials $V(h^+,H)$ and $V(k^{++},H)$ to determine the viable candidates for minimal models. In the subsequent sections we present the results for each case.
\section{Simplifying the Zee Model\label{sec:simple_zee}}
We have argued that any model of radiative neutrino masses, whose simplicity is to rival that of existing models, will likely be obtained by extending the SM to include either the charged singlet $h^+$ or the doubly charged singlet $k^{++}$, in conjunction with a second (as yet unspecified) field (call it $S$). In this section we search for a viable field $S$ when the singly charged singlet is added to the SM. 

The new field $S$ will appear in the scalar potential and must  provide a coupling that breaks lepton number symmetry.  Achieving this requires $S$ to appear in a Lagrangian term containing $h^+$ (which, in our conventions,  is assigned a nonzero lepton number due to its Yukawa coupling). To systematically search for candidate fields one can consider direct products of the form $h^m\otimes H^n$, where the integers obey $1\le m+n\le 3$.  A viable scenario is found if one such direct product contains a term with the quantum numbers $(1,R_S,-Y_S)\subset h^m\otimes H^n$, and the addition of the field $S\sim(1,R_S,Y_S)$ causes lepton number symmetry to be explicitly broken. Note that, in general, it is necessary for $S$ to appear explicitly in a term containing $h$, but this is not sufficient to ensure lepton number symmetry violation. To be sure the symmetry is broken, one typically requires $S$ to appear in two terms which make conflicting demands on the assignment of lepton number to $S$. Therefore one can search the direct products $h^m\otimes H^n$ for pairs of terms that have the same quantum numbers; if these terms require different lepton number symmetry assignments for $S$, the symmetry will be broken.

The direct products $h^m\otimes H^n$ contain the following terms with a single field type:
\begin{center}
\begin{tabular}{lll}
$\bullet \quad h\sim (1,2)$& \quad& $\bullet\quad H\sim (2,1)$ \\
\\
$\bullet \quad h^2 \sim (1,0)\oplus (1,4)$& \qquad\qquad\qquad\qquad&$\bullet \quad H^2\sim (3,2)\oplus (1,0)\oplus (3,0)$\\
\\
$\bullet \quad h^3\sim (1,2)\oplus (1,6)$& &$\bullet \quad H^3\sim (4,3)\oplus (4,1)\oplus (2,1)$
\end{tabular}
\end{center}
and the additional mixed terms:
\begin{center}
\begin{tabular}{l}
$\bullet \quad h\otimes H\sim (2,3)\oplus (2,1)$ \\
\\
 $\bullet \quad h^2\otimes H\sim (2,1)\oplus (2,5)\oplus (2,3)$\\
\\
$\bullet \quad h\otimes H^2\sim(3,4)\oplus (3,0)\oplus (1,2)\oplus (3,2)$
\end{tabular}
\end{center}
We only show the $SU(2)_L\otimes U(1)_Y$ quantum numbers in the above, and do not consider terms which differ only by the sign of hypercharge. Note that there is a slight abuse of notation here. For example, a given field operator can also stand for the charge- or hermitian-conjugate fields, so the $h^2$-term is symbolic for the following:
\bea
 h^2\sim (1,0)\oplus (1,4)& \longrightarrow& [h^-\otimes h^+\sim (1,0)] \quad \mathrm{and}\quad[h^+\otimes h^+\sim(1,4)].
\eea
The same is true for the other terms.

The above list, when combined with the fact that $\overline{e^c_R}\otimes e_R\sim(1,1,-4)$,  provides us with four candidates for the new scalar. For example, we observe that the $h\otimes H$ term and the $h^2\otimes H$ term both contain the representation $(1,2,3)$. Therefore if we add the doublet scalar $\Phi\sim(1,2,3)$ to the model, the new field appears in two terms that make conflicting demands on the lepton number assignment for $\Phi$, resulting in explicit lepton number violation.  The other candidates are the SM-like doublet  $(1,2,1)$, the doubly charged singlet $(1,1,4)$, and the real scalar triplet $(1,3,0)$. The first case is the original Zee model; lepton number symmetry can be successfully broken and radiative masses result. Similarly, the case with $(1,1,4)$ corresponds to the Zee-Babu model, which realizes mass at the two-loop level. Let us now discuss the two new candidates, each in turn.

\subsection{Model with a Real Scalar Triplet\label{subsec:triplet_zee}}
Upon extending the SM by the inclusion of the scalar singlet $h^+\sim(1,1,2)$ and the real scalar triplet $\Delta\sim(1,3,0)$, the Lagrangian contains the following terms
\bea
\mathcal{L}&\supset & f \,\overline{L^c}\,L\,h^+\ +\  \mu\, H^\dagger \Delta H \ +\ \lambda\, h^- \tilde{H}^\dagger \Delta H.\label{eq:triplet_new_lagrangian_terms}
\eea
which are sufficient to ensure that lepton number symmetry is explicitly broken. Note that $\Delta$ does not couple directly to SM fermions so the model requires less parameters than the Zee model.\footnote{For earlier works with the real scalar triplet see~\cite{Ross:1975fq} and references therein.} The second term plays an important role in models with the real triplet. An analysis of the full scalar potential shows that this term induces a VEV for $\Delta$ after electroweak symmetry breaking:
\bea
\langle \Delta \rangle&\simeq& \mu\, \frac{\langle H\rangle^2}{M_\Delta^2}.
\eea
Thus, in the limit $M_\Delta\gg \langle H\rangle$, this VEV is naturally suppressed relative to the weak scale. This is the same type of natural VEV suppression experienced by the triplet $T\sim(1,3,2)$ in the Type-II seesaw mechanism~\cite{type2_seesaw}, and its realization in the present model does not depend on the inclusion of $h^+$. The VEV of the scalar triplet is constrained by measurements of the $\rho$-parameter, as $\langle \Delta\rangle\ne0$ contributes to electroweak symmetry breaking and modifies the relationship between the $W$ and $Z$ boson masses. The corresponding bound is $\langle \Delta\rangle \lesssim\mathcal{O}(1)$~GeV~\cite{Nakamura:2010zzi}, which is readily obtained without tuning.
\begin{figure}[ttt]
\begin{center}
        \includegraphics[width = 0.5\textwidth]{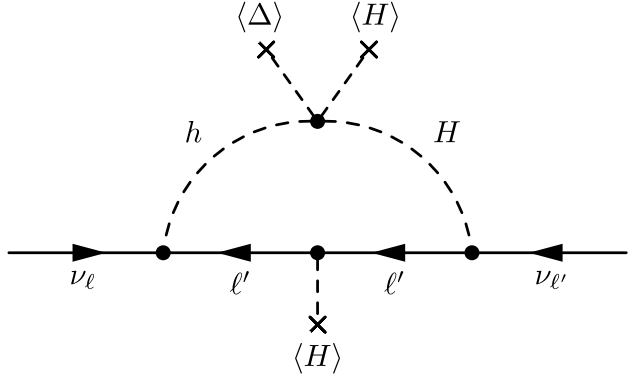}
\end{center}
\caption{A Zee-Like Model with a Real Scalar Triplet $\Delta\sim(1,3,0)$.}\label{fig:triplet_zee}
\end{figure}

\begin{figure}[h]
\begin{center}
        \includegraphics[width = 0.45\textwidth]{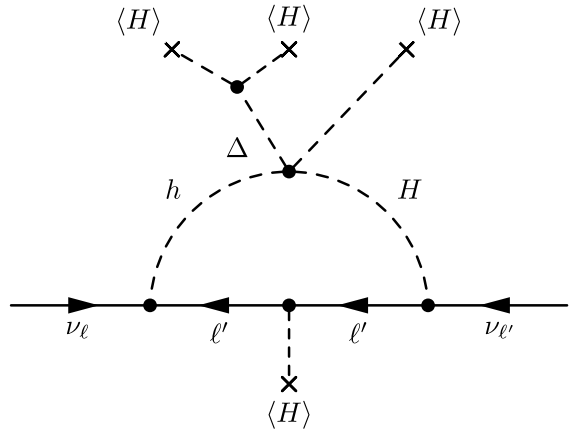}
\end{center}
\caption{The Explicit $d=7$ Diagram for the Triplet Model}\label{fig:triplet_zee_d7}
\end{figure}

Combining the Lagrangian terms in Eq.~\eqref{eq:triplet_new_lagrangian_terms} with those of the SM allows one to draw the one-loop diagram in Figure~\ref{fig:triplet_zee}. Thus, the model generates massive neutrinos. The one-loop  diagram is very similar to the original Zee model, except that the triplet VEV plays the role of the second SM-like doublet employed by Zee.  Note that neutrinos acquire mass via an effective operator with mass dimension $d=7$ in the low-energy theory; for completeness we demonstrate this in Figure~\ref{fig:triplet_zee_d7}. This is contrary to the Zee model, which realizes the Weinberg operator at low energies~\cite{Ma:1998dn}. 
\subsection{Model with a Charged Scalar Doublet\label{subsec:doublet_zee}}
\begin{figure}[t]
\begin{center}
        \includegraphics[width = 0.45\textwidth]{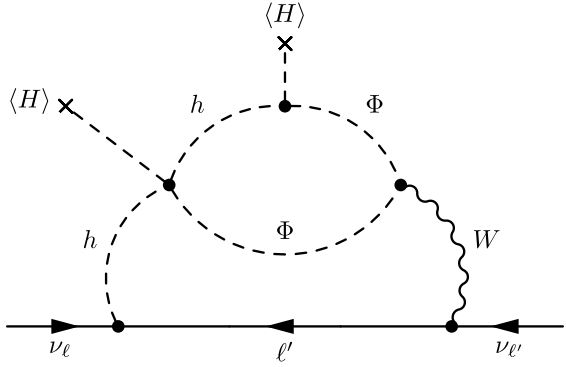}
\end{center}
\caption{A Two-Loop Model with a Charged Scalar Doublet $\Phi\sim(1,2,3)$.}\label{fig:doublet_zee}
\end{figure}

Let us now extend the SM to include the singlet $h^+\sim(1,1,2)$ and the charged doublet $\Phi=(\phi^{++},\phi^{+})^T\sim(1,2,3)$. The Lagrangian contains the following new terms:
\bea
\mathcal{L}&\supset& f \,\overline{L^c}\,L\,h^+\ +\  \lambda\, h^+h^+\Phi^\dagger\tilde{H}\ +\ \mu\, h^+\Phi^\dagger H,\label{eq:doublet_lagrangian}
\eea
which are again sufficient to ensure that lepton number symmetry is explicitly broken. Note that $\Phi$ does not couple directly to SM fermions so the model appears to be more minimal than the Zee model. Combining the Lagrangian terms \eqref{eq:doublet_lagrangian} with those of the SM leads to the mass diagram in Figure~\ref{fig:doublet_zee}. Thus, the model generates neutrino mass at the two-loop level. None the less, the model shares a number of features in common with the Zee model, as we will discuss in the next subsection.

Before ending this subsection, let us note that in radiative models of neutrino mass the gauge symmetries of the theory generally preclude a bare Lagrangian term that generates tree-level neutrino masses; therefore they also forbid an explicit counter-term to cancel any divergences that arise in loop diagrams that generate neutrino mass. The loop diagrams that give rise to neutrino mass in renormalizable radiative-mass models are therefore finite and calculable (though not always with ease).  This holds for the doublet model and the triplet model presented here, but also for radiative models more generally.
\subsection{Comparison with the Zee Model\label{subsec:zee_comparison}}
\begin{figure}[ttt]
\begin{center}
        \includegraphics[width = 0.5\textwidth]{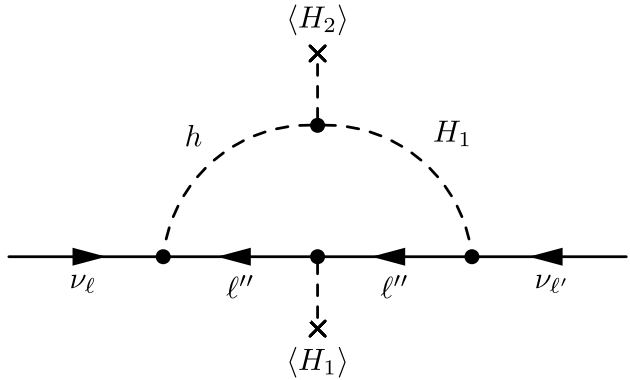}
\end{center}
\caption{The One-Loop Diagram of the  Zee Model.}\label{fig:zee}
\end{figure}

The Zee model extends the SM to include the charged singlet $h^+\sim(1,1,2)$ and an additional SM-like doublet [label the doublets as $H_{1,2}\sim(1,2,1)$]. The second doublet plays the crucial role of allowing the term $\mu_{\mathrm{Z}}\, h^- \tilde{H}_2^\dagger H_1\subset\mathcal{L}_{\mathrm{Zee}}$, without which lepton number symmetry would not be broken. Combining this term with the Yukawa coupling $f\,\overline{L^c}Lh^+\subset\mathcal{L}_{\mathrm{Zee}}$, leads to a one-loop diagram for neutrino masses, which, for completeness, we show in Figure~\ref{fig:zee}. 

Both doublets are required to develop nonzero VEVs and, in general, both couple to SM fermions (and contribute to fermion masses after symmetry breaking). The Yukawa couplings for $H_{1,2}$ are therefore non-diagonal in flavor space, increasing the number of parameters and leading to new flavor changing effects. It was long ago realized that a restricted version of the Zee model is obtained by imposing a discrete symmetry, such that only one doublet couples directly to SM fermions (the Zee-Wolfenstein model~\cite{Wolfenstein:1980sy}). In addition to removing the flavor changing effects and reducing the number of parameters, this forces $H_1$ to play the role of the SM doublet, while $H_2$ is relegated to the role of a spectator scalar that enables the one-loop diagram of Figure~\ref{fig:zee}.

After imposing this symmetry the Yukawa couplings for $H_1$ are diagonal, as in the SM. Consequently the external fermion $\nu_{\ell'}$ in Figure~\ref{fig:zee} must have the same flavor as the the internal charged lepton,  forcing $\ell''\rightarrow\ell'$. As the Yukawa couplings for $h^+$ are antisymmetric, $f\equiv f_{\ell\ell'}=-f_{\ell'\ell}$, the restricted model requires $\nu_\ell\ne\nu_{\ell'}$, or equivalently $\ell\ne\ell'$. The resulting mass matrix is therefore traceless, due to a vanishing diagonal.

Let us compare this with the triplet and doublet models. It is clear from Figure~\ref{fig:triplet_zee} that the triplet model bears a resemblance to the Zee model; the two one-loop diagrams have the same form, modulo the replacement $\mu_Z\rightarrow\lambda\langle H\rangle$ and $\langle H_2\rangle\rightarrow \langle \Delta\rangle$. However, whereas $H_2$ can  appear as a propagating degree of freedom inside the loop of Figure~\ref{fig:zee}, the triplet can only appear as an external leg in Figure~\ref{fig:triplet_zee}. The Yukawa couplings for $H$ are therefore diagonal in the triplet model, and the triplet itself is a spectator field that enables the lepton number symmetry violation necessary to realize Figure~\ref{fig:triplet_zee}. Thus, the triplet model is similar to the Zee-Wolfenstein model; it has a reduced number of parameters because only one scalar couples directly to SM fermions, and the two external neutrinos must have different flavors, ($\nu_\ell\ne\nu_{\ell'}$), giving a mass matrix with a vanishing diagonal. Different from the Zee-Wolfenstein model, however, these features arise without the imposition of an additional \emph{ad hoc} symmetry.

A similar conclusion  is reached for the doublet model of Figure~\ref{fig:doublet_zee}. Although Figure~\ref{fig:doublet_zee} looks somewhat different to the Zee diagram, the model shares some features with the Zee-Wolfenstein model. Specifically, in both models the loop diagram contains a single flavor-changing vertex, involving the charged singlet $h^+$, and a second flavor-conserving vertex. These features ensure that both models require $\nu_\ell\ne\nu_{\ell'}$, and the resulting mass matrices possess a vanishing diagonal. However, in the doublet model these features do not require an additional symmetry to be imposed. The doublet model also has less free parameters than Zee's original proposal.

We learn that both the triplet model and the doublet model are very similar to the Zee-Wolfenstein model, and in fact lead to neutrino mass matrices with the same texture. Modulo the need for the additional symmetry, the Zee-Wolfenstein model is generally considered to be simpler than the full Zee model. Its similarity to the doublet and triplet models supports our claim that the latter two are the simplest models of radiative neutrino masses. However, this similarity to the Zee-Wolfenstein model is also sufficient to exclude the doublet and triplet models as viable theories of neutrino mass. 
This is seen as follows. Due to the tracelessness of the mass matrix in the Zee-Wolfenstein model, the mass eigenvalues obey the sum condition $\sum_i m_i=0$. The neutrino mass eigenvalues are therefore completely determined by the experimentally observed mass-squared differences~\cite{He:2003nt}. By combining the atmospheric and solar data one can consequently show that the solar and atmospheric mixing angles cannot simultaneously lie within experimental limits~\cite{He:2003ih}. Thus, mass matrices with a vanishing diagonal are experimentally excluded. This conclusion allows one to exclude the triplet and doublet models of Sections~\ref{subsec:triplet_zee} and~\ref{subsec:doublet_zee}, respectively, without entering into the details of the model.\footnote{Note that higher order corrections will correct the mass matrix and generate diagonal elements. However, these effects are too small to modify our conclusions.}

Before concluding this section we note that the non-restricted Zee model, with two doublets that couple to SM fermions, cannot be excluded on the basis of the above reasoning~\cite{Balaji:2001ex}. Also, we point out that both the doublet model and the triplet model can be thought of as giving rise to an effective SM-like doublet that does not couple directly to SM fermions. In the doublet model one can think of $H'=h^-\times \Phi\sim (1,2,1)$ as playing the role of an effective SM-like doublet, while in the triplet model one has $H'=\Delta \times H\sim (1,2,1)$, where the triplet is written in $2\times2$ matrix form. This is why these models are so similar to the Zee-Wolfenstein model; they effectively generate the spectator doublet found in the Zee-Wolfenstein model.
\section{Models with a Doubly  Charged Singlet\label{sec:simple_zee_babu}}
In Section~\ref{sec:criteria} we came to the conclusion that the simplest models of radiative neutrino mass were likely to employ either the charged singlet $h^+\sim(1,1,2)$ or the doubly charged singlet $k^{++}\sim(1,1,4)$. We considered the singly charged scalar in the preceding section, finding two models that are essentially simplified versions of the Zee model. In this section we turn our attention to the doubly charged scalar.

We have followed the method of Section~\ref{sec:simple_zee} to systematically search for candidate fields that can be added to the SM, in addition to the doubly charged scalar, to realize radiative neutrino masses. Beyond the addition of the singly charged scalar $h^+$, which realizes the Zee-Babu model~\cite{Zee:1985id,Babu:1988ki,Babu:2002uu}, we did not find any additional viable candidates. Note that generalizations of the lepton number violating coupling $\mu k^{--}h^+h^+$ were considered in our analysis, with a new field $S\sim(1,R_S,2)$ added to produce the term $\mu k^{--}S S$. Here $R_S$ must be odd, and the case $R_S=1$ gives the Zee-Babu model. For $R_S=3$, lepton number violation occurs but $S$ is a triplet that couples directly to SM leptons, giving rise to tree-level masses (the Type-II seesaw). For $R_S=5$ there is no explicit lepton number violation in the model, though the lepton number symmetry is broken when $S$ acquires a VEV. This leads to a massless Goldstone mode (the Majoron), which we exclude.

There is one additional candidate scenario that is worth mentioning, as we will return to it in the following section. Consider adding the fields $k^{++}$ and $\Phi\sim(1,2,3)$ to the SM. Then it appears that one can have the following Lagrangian terms
\bea
\mathcal{L}&\supset& f\, \overline{e^c_R}\,e_R\,k^{++}\ +\ \mu\,k^{++} \Phi^\dagger\tilde{H}\ +\ \lambda\, (\Phi^\dagger\tau_a H)\cdot(\tilde{H}^\dagger\tau_aH)
\eea
which would be sufficient to break lepton number symmetry and generate neutrino masses. Here $\tau_a$ are the $SU(2)$ generators, and the last term is the product of two $SU(2)$ triplets. However, one can show that the last term vanishes identically because the two triplets  are constructed with three occurrences of the single doublet $H$. This scenario is therefore not viable and we conclude that the Zee-Babu model appears to the simplest model of radiative neutrino mass that employs the doubly charged singlet.
\section{Allowing Additional Complexity\label{sec:more_complexity}}

One can ask what models are possible if further complexity, in the form of a third beyond-SM field,  is permitted. Models with three beyond-SM fields certainly exist, and some examples have been discussed in Ref.~\cite{FileviezPerez:2009ud}. At this level of complexity one can realize radiative masses with new colored fields without having to impose a symmetry. For example, one can extend the SM to include the $SU(3)_c$ scalar octet $\Sigma\sim(8,2,1)$ and two adjoint fermions $\rho_{1,2}\sim(8,3,0)$, or alternatively one can employ $\Sigma$ with two copies of the fermion $\rho_{1,2}\sim(8,1,0)$, and achieve viable neutrino masses at the one-loop level~\cite{FileviezPerez:2009ud}.\footnote{Variations  are possible with two copies of $\Sigma$  and a single fermion $\rho$, in both of these cases~\cite{FileviezPerez:2009ud}.} The model of Ref.~\cite{Kumericki:2012bh} also achieves radiative one-loop masses with the scalar $S\sim(1,4,1)$ and at least two quintuplet fermions $\mathcal{F}_R\sim(1,5,0)$, for large values of $M_\mathcal{F}$. More generally, a class of radiative models is obtained when the new physics is heavy in the tree-level seesaws discussed in Ref.~\cite{McDonald:2013kca}.

\begin{figure}[ttt]
\begin{center}
        \includegraphics[width = 0.5\textwidth]{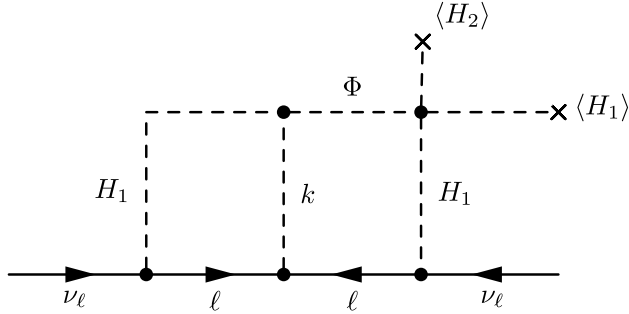}
\end{center}
\caption{Two-Loop Diagram for Neutrino Masses in the Model with a Doubly Charged Singlet and a Charged Doublet.}\label{fig:doublet_2loop}
\end{figure}

We will not attempt to systematically determine all viable candidates in this case, as the possibilities appear to be more numerous. However, we will point out one model which is readily inferred from the previous discussion and, to the best of our knowledge, has not appeared in the literature. Based on the discussion in Section~\ref{sec:simple_zee_babu}, it should be clear that a model of radiative masses can be obtained by extending the SM to include the doubly charged singlet $k^{++}\sim(1,1,4)$, the charged doublet $\Phi\sim(1,2,3)$ and an extra SM-like Higgs doublet [denote the pair as $H_{1,2}\sim(1,2,1)$]. This model has the following Lagrangian terms:
\bea
\mathcal{L}&\supset& f\, \overline{e^c_R}\,e_R\,k^{++}\ +\  \mu_i\,k^{++} \Phi^\dagger\tilde{H}_i\ +\ \lambda_{ij}\, (\Phi^\dagger\tau_a H_i)\cdot(\tilde{H}_j^\dagger\tau_aH_j)\,,
\eea
which are sufficient to break lepton number symmetry and realize radiative masses at the two-loop level, as shown in Figure~\ref{fig:doublet_2loop}. For simplicity we have used a single flavor label (``$\ell$") for all four fermion lines in the figure, though the fermions can have different flavors. Relative to the models in the preceding sections, this scenario has an additional field and requires more parameters, due partly to the non-diagonal Yukawa couplings for the SM-like doublets.
 However, we note that a restricted version of this model exists, \emph{\`a la} Wolfenstein, such that only one of the doublets couples to SM fermions, while the second doublet is a spectator 
field that enables lepton number symmetry breaking. We postpone a detailed discussion of this model, and the phenomenology of the charged doublet, for a future work. However, its existence exemplifies the greater freedom found in more complex models.

\begin{figure}[ttt]
\begin{center}
        \includegraphics[width = 0.5\textwidth]{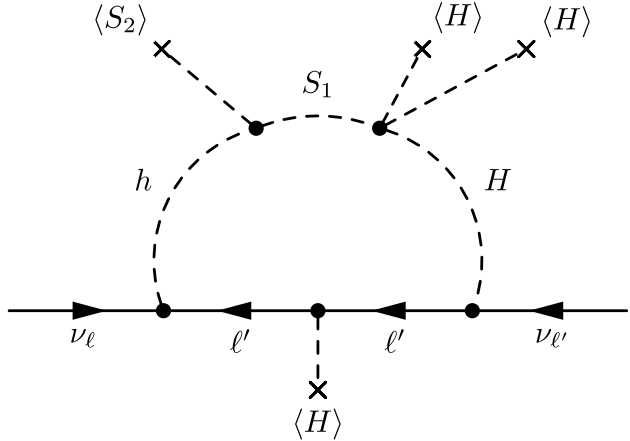}
\end{center}
\caption{One-Loop Diagram for Neutrino Mass in a Variant Zee Model Containing the Quadruplet Scalars $S_{1,2}\sim(1,4,1)$.}\label{fig:complicated_zee}
\end{figure}

We also note that there is an obvious generalization of the Zee model that arises with three new multiplets. The Zee model employs the coupling $\mu_\mathrm{Z} h^-\tilde{H}^\dagger_2H_1\subset\mathcal{L}_{\mathrm{Zee}}$ to break lepton number symmetry. One can generalize this by considering models containing the term $\mu h^- \tilde{S}^\dagger_2S_1$, with the scalars $S_{1,2}\sim(1,R_S,1)$. Note that $R_S$ must be even to ensure that $S_{1,2}$ contain a neutral component, and two fields are required because the $SU(2)$ singlet obtained from $\tilde{S}_1^\dagger S_1$ vanishes whenever $R_S$ is even (this is the generalization of the vanishing of the $\mu_\mathrm{Z}$-term when only one doublet is employed in the Zee model).  The choice $R_S=2$ gives the Zee model, while for $R_S=4$ the model has the following Lagrangian terms 
\bea
\mathcal{L}&\supset& f \,\overline{L^c}\,L\,h^+\ +\  \lambda_i\, S_i \tilde{H}H^\dagger H\ +\ \mu\, h^-\tilde{S_2}^\dagger S_1.\label{eq:complex_zee}
\eea
These ensure that lepton number symmetry is broken, leading to neutrino masses at the one-loop level, as shown in Figure~\ref{fig:complicated_zee}. The term linear in $S_i$ is essential to ensure the symmetry is broken. For even-valued  $R_S>4$ such a term is not possible and radiative neutrino masses do not arise.\footnote{We do not consider models where global lepton number symmetry is an exact symmetry that is spontaneously broken, due to the resulting Majoron.} Thus only $R_S=2$ and $R_S=4$ lead to radiative masses. A generalization of the Wolfenstein symmetry is also possible, such that the $S_2H^3$ term is forbidden and $S_2$ is a spectator field, with only $S_1$ appearing in the loop. As with the colored one-loop models of Ref.~\cite{FileviezPerez:2009ud}, two of the new multiplets have the same quantum numbers in this framework. However, as with the simplified Zee models, the mass matrix can again be excluded due to a vanishing diagonal.

In excluding the doublet model, the triplet model, and the aforementioned model with the quadruplet scalars $S_{1,2}\sim(1,4,1)$, one observes a common theme that bears mentioning.  Each of these models leads to a neutrino mass matrix with a vanishing diagonal, producing a spectrum of massive neutrinos that is now experimentally excluded. The form of the mass matrices can be traced back to the fact that the loop diagrams contain a single flavor changing vertex, involving the charged singlet $h^+$, which is attached to the virtual charged-lepton line. The internal charged lepton is re-converted to a neutrino by a flavor conserving interaction with a SM (or SM-like) doublet and, as the charged-singlet couplings $f_{\ell\ell'}$ are anti-symmetric, the resulting mass matrix has a vanishing diagonal.\footnote{In the doublet model the flavor conserving vertex involves a $W$ boson.} These features are not sensitive to the origin of the coupling between the charged singlet and the SM doublet; any model that gives rise to neutrino mass via a loop-diagram with these features will be excluded, regardless of the details of the beyond-SM physics that couples $h^+$ and $H$. The same conclusions are drawn for models employing $h^+$ and a $W$ boson at the vertices. The three models we have detailed are the simplest examples, but more complicated models that retain these features will also be excluded. This is a relatively powerful statement, and it enables one to rule out a class of models \emph{purely on the basis of neutrino mass and mixing data}. One does not have to utilize any other constraints or direct bounds to rule out such models. This truly demonstrates the power of recent advances in experimental neutrino physics. 

\begin{figure}[ttt]
\begin{center}
        \includegraphics[width = 0.5\textwidth]{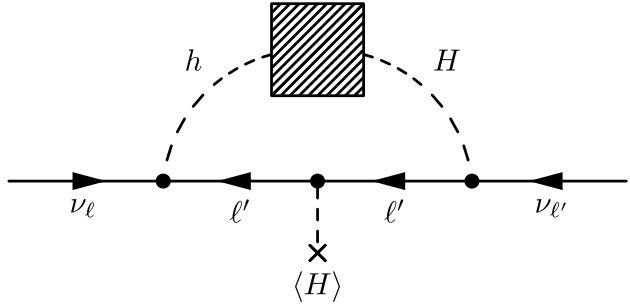}
\end{center}
\caption{Schematic for a Class of One-Loop Diagrams that are Experimentally Excluded.}\label{fig:general_zee}
\end{figure}

The general mass diagram for this class of model is given in Figure~\ref{fig:general_zee}, with a similar diagram possible when the internal SM doublet $H$ is replaced by a $W$ boson. The hatched box contains unspecified beyond-SM physics that gives rise to an effective coupling between $h^+$ and the SM doublet; regardless of the details of this unspecified physics, the resulting model can be excluded. 


\section{Discussion and Conclusion\label{sec:conc}}
Adopting the view that the imposition of \emph{ad hoc} symmetries is the least desirable approach, the simplicity of a given  model of radiative neutrino mass can be assessed by its particle content and the number of free parameters it introduces. In this work we searched for the simplest models of radiative neutrino mass, concluding that these have two beyond-SM fields. In addition to the well known Zee and Zee-Babu models, we found two new models that utilize the charged singlet $h^+\sim(1,1,2)$; namely the triplet model, with $\Delta\sim(1,3,0)$, and the (charged) doublet model, with $\Phi\sim(1,2,3)$. Both of these models are, in some sense, simplified versions of the original Zee model, and they appear to be the simplest models of radiative neutrino masses. We find that these four models are the only radiative mass models with just two beyond-SM multiplets that are devoid of \emph{ad hoc} new symmetries. However, despite their simplicity, or perhaps because of it, the new models can be excluded. This leaves the original models of Zee and Zee-Babu as the remaining \emph{viable} models of radiative masses with just two new multiplets --- a result that somehow seems  fitting. 

What conclusions can be drawn upon having uncovered these simple models, and having excluded them? Based on simplicity arguments alone, many would suspect that neutrino masses are generated by a high energy seesaw via the exchange of heavy gauge-singlet neutrinos~\cite{type1_seesaw}. Despite its simplicity, however, such a scenario is incredibly difficult to verify. If our best hope is to simply exclude alternatives to this (Type-I) seesaw, one can consider the present work as a step in this direction. 

More generally one can ask what these results may teach us about radiative models of neutrino mass. If one is inclined to assume that the simplest models are more probable and/or desirable, then the exclusion of the triplet and doublet models may be considered as \emph{circumstantial} evidence that radiative neutrino masses are not realized in Nature.  The exclusion of the Zee model itself would provide further support for this train of thought. Of course, ruling out the minimal cases does not exclude radiative neutrino masses, as models with additional complexity certainly exist; indeed, we presented a new example in Section~\ref{sec:more_complexity}. It does, however, provide food for thought.

In addition to excluding the doublet and triplet models, we showed that a generalized Zee model containing the scalars $S_{1,2}\sim(1,4,1)$ is also excluded. Collectively these models reveal a common theme; there is in fact a more general class of models that can be excluded purely on the basis of neutrino mass and mixing data. A schematic for the loop diagram in these models appears in  Figure~\ref{fig:general_zee}, and similar models where the $W$ boson plays the (flavor conserving)  role of $H$ are also excluded. These observations show the power of available mass and mixing data to differentiate between candidate models of radiative neutrino mass in a meaningful way. 

Finally we note that, although we have ruled out the triplet and doublet models, it may be that the embedding of these frameworks in more-complete UV theories will introduce additional ingredients. For example, a viable model with a  neutrino mass matrix whose diagonal vanishes can be obtained within an $SO(10)$ Grand Unified Theory~\cite{Grimus:2008tm}, due to additional mixing effects from the  charged lepton sector. Also, it is known that a light triplet can arise in non-supersymmetric Grand Unified Theories~\cite{Dorsner:2005fq}, and one wonders if the present observations could impact the origin of neutrino mass in such models. Similarly, models with a flavor symmetry may contain additional ingredients, just as the implementation of an $A_4$ flavor symmetry in the Zee model requires additional fields~\cite{Fukuyama:2010ff}. These matters may deserve further attention.

\section*{Acknowledgments\label{sec:ackn}}
The authors thank A.~G.~Akeroyd, C.~H.~Chen, A.~Kobakhidze, T.~Schwetz, and R.~R.~Volkas. SSCL is supported in part by the NSC under Grant No.
NSC-101-2811-M-006-015 and in part by the NCTS of Taiwan. KM is supported by the Australian Research Council.
\appendix
\section{Models with Colored Fields}
Ref.~\cite{FileviezPerez:2009ud} also endeavored to determine the simplest models of radiative neutrino masses. However, they arrived at different conclusions to us and, in particular, have not found the triplet and doublet models detailed in this work. We briefly explain why. Although their method of judging simplicity is similar to ours (e.g.~they exclude models that require one to impose baryon number symmetry), Ref.~\cite{FileviezPerez:2009ud} gave preference to models with new colored fields due to their prospects of discovery at the LHC. However, as it turns out, the demands of absolute simplicity and new colored fields are not concordant. Thus, they arrive at new one-loop models with three new colored fields, a subset of which are viable, and argue that these are the simplest one-loop models beyond the original Zee proposal. This is contrary to the results presented here.   Their focus on one-loop models (obviously) ensures they do not consider the two-loop model of Section~\ref{sec:more_complexity}, which also has three new fields.

\end{document}